# Spaser operation below threshold: autonomous vs. driven spasers

E. S. Andrianov,[1,2] A. A. Pukhov,[1,2,3] A. V. Dorofeenko,[1,2,3] A. P. Vinogradov,[1,2,3] and A. A. Lisyansky[4,5]

[1]*Moscow Institute of Physics and Technology, 9 Institutskiy per., Dolgoprudniy 141700, Moscow Reg., Russia*
[2]*All-Russia Research Institute of Automatics, 22 Sushchevskaya, Moscow 127055, Russia*
[3]*Institute for Theoretical and Applied Electromagnetics RAS, 13 Izhorskaya, Moscow 125412, Russia*
[4]*Department of Physics, Queens College of the City University of New York, Flushing, NY 11367, USA*
[5]*The Graduate Center of the City University of New York, New York, New York 10016, USA*

**Abstract:** At the plasmon resonance, high Joule losses in a metal nanoparticle of a spaser result in its low *Q*-factor. Due to the latter, to achieve the spasing regime, in which the number of coherent plasmons exceeds the number of incoherent plasmons, unsustainably high pump rates may be required. We show that under the condition of loss compensation by a spaser driven by an external optical wave, the number of coherent plasmons increases dramatically, and the quantum noise is suppressed. Since the compensation of losses of the driving wave may occur even near the spasing threshold, the number of coherent plasmons may exceed the number of spontaneously excited plasmons at achievable pump rates.

## 1. INTRODUCTION

In the last decade, the plasmonic analog of the laser, the spaser [1], has attracted much attention [2-8]. Although unambiguous experimental realizations of a single 3-D confined subwavelength spaser have not yet been accomplished, the physics of spasers is very rich and potential applications motivate continuing studies. In Refs. [9,10] it was shown that near the spaser generation threshold, the number of coherent plasmons is much smaller than the number of incoherent plasmons. Therefore, to achieve a signal-to-noise ratio greater than unity, a pump rate much greater than the threshold value is required. For conventional semiconductor active media, such a pump rate would lead to destruction of the spaser [9-12]. A possible solution for this situation could be an increase in the efficiency of an active medium, as was suggested in Ref. [13], in which the use of ballistic quantum wires with low-levels of quantum fluctuations was proposed.

It seems, however, that a single 3-D confined autonomously generating spaser may only have a limited use [3,9,10,13]. The use of a spaser array distributed in metamaterials may have much better perspective [14-19]. Metamaterials are expected to have numerous promising applications in all-optical and optoelectronic devices that have been broadly discussed [20-22]. However, as of today, the suggested applications of metamaterials cannot be realized due to high Joule losses. To overcome this problem and compensate for losses it was proposed to incorporate an active medium into the material matrix [23-26]. Since all known metamaterials contain metal elements, active molecules coupled to the metal elements may transform the system into a composite filled with spasers. In this case, each spaser is subjected to an external field, which is either the field of neighboring spasers or an external field. The coherent component of oscillations of the spaser dipole moment is determined not only by the generation of the respective mode due to incoherent pumping but also by the moment induced by the external field. Obviously, increasing the external field amplitude would result in an increase of the signal-to-noise ratio. However, this may also increase losses in the system. Thus, loss compensation along with suppressed noise may not be possible to achieve by simply increasing the amplitude of the external field. Currently, the question whether this desirable situation can be realized remains opened.

It was shown that an exact compensation of both the work of the external field synchronizing the spaser and Joule loss of this field in the spaser can be attained for a pump rate below the threshold [27,28]. Thus, a pump rate insufficient for undamped self-oscillations of an autonomous spaser can be adequate for both maintaining undamped plasmon oscillations and compensating for Joule loss of an external field in the driven spaser. This happens due to a decrease of the amplitude of plasmon field oscillations so that their own Joule losses become smaller. The compensation can be observed in an external field with an amplitude that depends on the pumping rate and the frequency.

In Ref. [27], however, the spontaneous generation of plasmons was ignored. It is only natural to expect that below the threshold, the power of the spontaneous generation would substantially exceed the power of the generation of coherent plasmons initiated by an external wave, due to the smallness of the wave amplitude required for loss compensation. Thus, there is a risk that the signal-to-noise ratio would be much smaller than unity similar to the autonomous spaser near the generation threshold. This would make spasers completely unusable for loss compensation. In this paper, we show that this is not so.

We obtain the conditions at which both the regime for loss compensation and the prevalence of undamped *coherent* plasmon oscillations over incoherent oscillations are realized. As a result, a wave may propagate in a metamaterial without attenuation because it is being fed coherently by spasers using the energy of incoherent pumping.

## 2. HAMILTONIAN OF THE DRIVEN SPASER

To introduce notations and remind the main equations and approaches describing spaser dynamics, we first review the dynamics of a spaser under the influence of an external optical field [29]. For the sake of generality, we model an active medium of a spaser as a two-level system (TLS). In reality, active media are multilevel. Among those levels are two working levels for which the frequency coincides with the frequency of the resonator mode. As a rule, the transition rate to the working levels is much greater than the relaxation rate from the working levels to other levels [30]. Therefore, one can treat an active medium as a TLS with an effective population inversion and an effective pumping rate [31]. Specifically for spasers it was shown that using three- and four-level systems does not produce new effects [5,32]. We assume that the distance between the plasmonic nanoparticle (NP) and the TLS is much smaller than the photon wavelength in vacuum so that these interact via the near field dipole interaction. We also assume that the TLS transition frequency, $\omega_{TLS}$, coincides with the plasmon resonance of the dipole mode of the NP, $\omega_{SP}$. The dynamics of the coupled TLS and NP in the external field can be described by the model Hamiltonian [33,34]

$$\hat{H} = \hat{H}_{NP} + \hat{H}_{TLS} + \hat{V}_{TLS-NP} + \hat{V}_{TLS-w} + \hat{V}_{NP-w}, \tag{1}$$

where $\hat{H}_{NP} = \hbar\omega_{SP}\hat{a}^\dagger\hat{a}$ and $\hat{H}_{TLS} = \hbar\omega_{TLS}\hat{\sigma}^\dagger\hat{\sigma}$ are Hamiltonians of NP's surface plasmons and the TLS, respectively, $\hat{a}$ and $\hat{a}^\dagger$ are plasmon annihilation and creation operators, $\hat{\sigma}$ is the transition operator between ground $|g\rangle$ and excited $|e\rangle$ states of the TLS. The operator $\hat{V}_{TLS-NP} = -\hat{\mathbf{d}}_{TLS}\cdot\mathbf{E}_{NP} = \hbar\Omega_R\left(\hat{a}^+\hat{\sigma} + \hat{\sigma}^+\hat{a}\right)$ describes the interaction of the TLS dipole moment $\hat{\mathbf{d}}_{TLS} = \boldsymbol{\mu}_{TLS}(\hat{\sigma} + \hat{\sigma}^\dagger)$ with the NP, $\boldsymbol{\mu}_{TLS} = \langle e|e\mathbf{r}|g\rangle$ is the TLS dipole moment matrix element, and $\hat{V}_{TLS-w} = -\hat{\mathbf{d}}_{TLS}\cdot\mathbf{E}$ and $\hat{V}_{NP-w} = -\hat{\mathbf{d}}_{NP}\cdot\mathbf{E}$ determine interactions of the TLS and NP dipole moments with the field $\mathbf{E}$ of the external wave, respectively, where $\hat{\mathbf{d}}_{NP} = \boldsymbol{\mu}_{NP}(\hat{a} + \hat{a}^\dagger)$. Denoting $-\mathbf{E}\cdot\boldsymbol{\mu}_{TLS}/\hbar$ as $\Omega$, we have $\hat{V}_{TLS-w} = \hbar\Omega(\hat{\sigma} + \hat{\sigma}^\dagger)$. We assume that the dipole moments of the TLS and the NP are collinear, then the operator $\hat{V}_{TLS-w}$ takes the form [34]

$$\hat{V}_{NP-w} = \alpha\hbar\Omega(\hat{a}^+ + \hat{a}), \tag{2}$$

where $\alpha = |\boldsymbol{\mu}_{NP}|/|\boldsymbol{\mu}_{TLS}|$.

Using the Heisenberg approach one can obtain a system of equations governing the time evolution of the operators. By doing this, we arrive at the well-known Maxwell-Bloch equations [1,2,4,34]

$$\dot{\hat{D}} = 2i\Omega_R(\hat{a}^\dagger\hat{\sigma} - \hat{\sigma}^\dagger\hat{a}) + 2i\Omega(\hat{\sigma} - \hat{\sigma}^\dagger) - \gamma_D\left(\hat{D} - \hat{D}_0\right) + \hat{F}_D(t), \tag{3}$$

$$\dot{\hat{\sigma}} = (i\delta - \gamma_\sigma)\hat{\sigma} + i\Omega_R\hat{a}\hat{D} + i\Omega\hat{D} + \hat{F}_\sigma(t), \tag{4}$$

$$\dot{\hat{a}} = (i\Delta - \gamma_a)\hat{a} - i\Omega_R\hat{\sigma} - i\alpha\Omega + \hat{F}_a(t), \tag{5}$$

where the operator $\hat{D} = [\hat{\sigma}^\dagger, \hat{\sigma}] = \hat{n}_e - \hat{n}_g$, describes the TLS population inversion, $\delta = \omega_a - \omega_{TLS}$ and $\Delta = \omega_a - \omega_{SP}$ are frequency detunings of the TLS transition and the surface plasmons excitation from the frequency of spaser auto-oscillation, and the operator $\hat{D}_0$ characterizes the magnitude of pumping [30,35]. We assume that $\omega_{SP} \approx \omega_{TLS} \approx \omega_a$. In Eqs. (3)-(5), following Refs. [1,2,36,37] dissipative and pumping terms are introduced phenomenologically by incorporating relaxation rates, $\gamma_\sigma$, and $\gamma_a$, for the respective quantities. The noise terms $\hat{F}_\sigma(t), \hat{F}_a(t)$, and $\hat{F}_D(t)$ with zero mean value have been introduced according to the fluctuation dissipative theorem [30]. The presence of dissipative and noise terms indicates an interaction with a reservoir. In the case of spasers, there are two reservoirs: photons in free space and thermal phonons in the metal NP and the TLS. We assume that the main loss in the system is due to Joule loss in the metal. Note that radiation losses can be neglected for small particles only [38]. The critical size of the particle depends on its material. For gold, the critical size is ~50 nm [39], for silver it is ~30-40 nm [40]. For NPs of sizes greater than 50 nm, radiation and Joule losses are comparable. However, even in this case system of equations (3)-(5) does not change. Thus, loss due to radiation only affects the constant $\gamma_a$ and does not change the character of the attenuation.

In Eqs. (3)-(5), plasmon annihilation and TLS transition operators are represented as $\hat{a}(t)\exp(-i\omega_a t)$ and $\hat{\sigma}(t)\exp(-i\omega_a t)$, respectively, where $\hat{a}(t)$ and $\hat{\sigma}(t)$ are slowly varying amplitudes. We use the rotating wave approximation [35], in which fast-oscillating terms proportional to $\exp(\pm 2i\omega_a t)$ are neglected. Solving an operator system of equations (3)-(5) is a complicated task. The usual approach involves replacing operators by $c$-numbers, $a(t)$, $\sigma(t)$ and $D(t)$, and uncoupling their correlators [3]. However, in this procedure, information of the incoherent component of surface plasmons arising due to spontaneous generation is lost.

In order to take into account incoherent plasmons, instead of replacing operators by $c$-numbers, one can move from operator equations (3)-(5) to equations for expected values retaining all correlations [29,37].

$$\left\langle \frac{d}{dt}\hat{D} \right\rangle = 2i\Omega_R(\langle \hat{a}^\dagger \hat{\sigma}\rangle - \langle \hat{\sigma}^\dagger \hat{a}\rangle) + 2i\Omega(\langle \hat{\sigma}\rangle - \langle \hat{\sigma}^\dagger \rangle) - \gamma_D\left(\langle \hat{D}\rangle - \langle \hat{D}_0\rangle\right), \qquad (3a)$$

$$\left\langle \frac{d}{dt}\hat{\sigma} \right\rangle = (i\delta - \gamma_\sigma)\langle \hat{\sigma}\rangle + i\Omega_R \langle \hat{a}\hat{D}\rangle + i\Omega\langle \hat{D}\rangle, \qquad (4a)$$

$$\left\langle \frac{d}{dt}\hat{a} \right\rangle = (i\Delta - \gamma_a)\langle \hat{a}\rangle - i\Omega_R\langle \hat{\sigma}\rangle - i\alpha\Omega, \qquad (5a)$$

In contrast to the semiclassical equations (3)-(5), these are exact equations correctly describing the Purcell effect. However, the new system of equations contains averages of products of operators ($\langle \hat{a}^\dagger \hat{\sigma}\rangle$, $\langle \hat{\sigma}^\dagger \hat{a}\rangle$, $\langle \hat{a}\hat{D}\rangle$) for which we need additional equations. Using the Heisenberg approach we can easily obtain the corresponding operator equations and transition to the equations for expected values that contain new averages of products of operators (e.g., $\hat{a}^\dagger \hat{\sigma}\hat{a}$, and $\hat{a}^\dagger \hat{\sigma}^\dagger \hat{a}$). This results in an infinite chain of equations. Below we consider spaser behavior near or below the threshold at which spasing begins. We call this the first spasing threshold, $D_{0th}^{(1)}$, as opposed to the second threshold, $D_{0th}^{(2)}$, at which the number of coherent plasmons becomes of the same order as the number of incoherent plasmons. Near $D_{th}^{(1)}$, due to large Joule losses in the NP ($\gamma_a \sim 10^{14} s^{-1}$ and $\Omega_R \ll \gamma_a$), the number of the excited plasmons is small, $\langle \hat{a}^\dagger \hat{a}\rangle \sim (\alpha\Omega/\gamma_a)^2 \leq 1$. Thus, we may limit our consideration to the Hilbert space containing states without plasmons and states with one plasmon. In this approximation, the expected values of operators containing two or more plasmon creation (annihilation) operators (e.g., $\langle \hat{\sigma}^+ \hat{D}\hat{a}^2\rangle$ or $\langle \hat{a}^+ \hat{D} \hat{a}^2 \rangle$) are neglected [34]. As a result, we obtain a closed system of equations for expected values of the operators $\hat{a}$, $\hat{a}^\dagger$, $\hat{\sigma}$, $\hat{\sigma}^\dagger$, $\hat{a}\hat{D}$, $\hat{a}^\dagger\hat{D}$, $\hat{a}^\dagger\hat{\sigma}\hat{a}$, $\hat{a}^\dagger\hat{\sigma}^\dagger\hat{a}$, $\hat{a}^\dagger\hat{a}$, $\hat{D}$, $\hat{a}^\dagger\hat{\sigma}$, $\hat{\sigma}^\dagger\hat{a}$, and $\hat{a}^\dagger\hat{D}\hat{a}$:

$$\left\langle \frac{d}{dt}\hat{a} \right\rangle = (i\Delta - \gamma_a)\langle\hat{a}\rangle - i\Omega_R \langle\hat{\sigma}\rangle - i\alpha\Omega,$$

$$\left\langle \frac{d}{dt}\hat{\sigma} \right\rangle = (i\delta - \gamma_\sigma)\langle\hat{\sigma}\rangle + i\Omega_R \langle\hat{a}\hat{D}\rangle + i\Omega\langle\hat{D}\rangle,$$

$$\left\langle \frac{d}{dt}(\hat{a}\hat{D}) \right\rangle = \gamma_D D_0 \langle\hat{a}\rangle + i\Omega_R \langle\hat{\sigma}\rangle + (-i\Delta - \gamma_a - \gamma_D)\langle\hat{a}\hat{D}\rangle$$
$$+ 2i\Omega_R \langle\hat{a}^+\hat{\sigma}\hat{a}\rangle - i\alpha\Omega\langle\hat{D}\rangle - i\Omega\langle\hat{\sigma}^+\hat{a}\rangle,$$

$$\left\langle \frac{d}{dt}(\hat{a}^+\hat{\sigma}\hat{a}) \right\rangle = i\Omega\langle\hat{a}\rangle/2 + i\Omega\langle\hat{a}\hat{D}\rangle/2 + (i\delta - 2\gamma_a - \gamma_\sigma)\langle\hat{a}^+\hat{\sigma}\hat{a}\rangle$$
$$- i\alpha\Omega\langle\hat{a}^+\hat{\sigma}\rangle - i\Omega\langle\hat{a}^+\hat{D}\hat{a}\rangle,$$

$$\left\langle \frac{d}{dt}(\hat{a}^+\hat{a}) \right\rangle = i\alpha\Omega(\langle\hat{a}\rangle - \langle\hat{a}^+\rangle) - 2\gamma_a\langle\hat{a}^+\hat{a}\rangle - i\Omega(\langle\hat{a}^+\hat{\sigma}\rangle - \langle\hat{\sigma}^+\hat{a}\rangle),$$

$$\left\langle \frac{d}{dt}\hat{D} \right\rangle = 2i\Omega(\langle\hat{\sigma}\rangle - \langle\hat{\sigma}^+\rangle) - \gamma_D\langle\hat{D}\rangle + 2i\Omega(\langle\hat{a}^+\hat{\sigma}\rangle - \langle\hat{\sigma}^+\hat{a}\rangle) + \gamma_D D_0,$$

$$\left\langle \frac{d}{dt}(\hat{a}^+\hat{\sigma}) \right\rangle = i\alpha\Omega\langle\sigma\rangle + i\Omega\langle\hat{a}^+\hat{D}\rangle + i\Omega\langle\hat{D}\rangle/2$$
$$+ (i(\delta - \Delta) - \gamma_a - \gamma_\sigma)\langle\hat{a}^+\hat{\sigma}\rangle + i\Omega\langle\hat{a}^+\hat{D}\hat{a}\rangle + i\Omega/2,$$

$$\left\langle \frac{d}{dt}(\hat{a}^+\hat{D}\hat{a}) \right\rangle = i\alpha\Omega(\langle\hat{a}\hat{D}\rangle - \langle\hat{a}^+\hat{D}\rangle) + 2i\Omega(\langle\hat{a}^+\hat{\sigma}\hat{a}\rangle - \langle\hat{a}^+\hat{\sigma}^+\hat{a}\rangle)$$
$$+ \gamma_D D_0 \langle\hat{a}^+\hat{a}\rangle + i\Omega(\langle\hat{a}^+\hat{\sigma}\rangle - \langle\hat{\sigma}^+\hat{a}\rangle) - (2\gamma_a + \gamma_D)\langle\hat{a}^+\hat{D}\hat{a}\rangle.$$

(6)

This system allows one to analyze the response of the spaser dipole moment to the external field in the low quantum regime. In particular, $|\langle\hat{a}\rangle|^2$ describes the energy of coherent oscillations, whereas $\langle\hat{a}^+\hat{a}\rangle$ describes the total energy of oscillations. Note that the obtained system takes into account both dissipation, via the attenuation constants, $\gamma_D, \gamma_\sigma$, and $\gamma_a$, and noise, via operator correlators.

For further calculation we use the parameter values that are usual for spasers [9,10,38]: $\gamma_a = 10^{14} s^{-1}$ is the value of the damping constant of plasmons in the NP, $\gamma_\sigma = 10^{11} s^{-1}$ is the dissipation rate of the TLS, the Rabi constant of the interaction between dipole moments of the NP and the TLS is $\Omega_R = 10^{13} s^{-1}$. The value of the pumping rate $\gamma_D$ is related to the inversion $D_0$ by the relation $\gamma_D = 2\gamma_\sigma(1+D_0)/(1-D_0)$ (see below). Experimental realizations of different values of these parameters is discussed in Conclusions.

## 3. SPASER DRIVEN BY AN EXTERNAL WAVE

The spaser interaction with an external field is rather complex. Without the external field, an autonomous spaser generates surface plasmons oscillating with a frequency determined by properties of the NP and the TLC. These oscillations can be synchronized by a sufficiently strong external field [29]. The region of the synchronization, the Arnold tongue, is defined by the pump power, the frequency detuning, and the external field intensity. In particular, below the spasing threshold the synchronization always happens.

We consider values of parameters that belong to the Arnold tongue, so that the external field synchronizes a spaser and it oscillates with the frequency $\omega_E$ of the driving field [29,41]. The work done by the external field on the spaser during one period is proportional to $E(\mu_{NP}\langle \hat{a} \rangle + \mu_{TLS}\langle \hat{\sigma} \rangle)$ [27,42]. The phase difference, $\Delta\varphi$, between the spaser dipole moment, $\mu_{NP}\langle \hat{a} \rangle + \mu_{TLS}\langle \hat{\sigma} \rangle$, and the external field defines the direction of the energy transfer [43]. When $0 < \Delta\varphi < \pi$, the field loses energy due to the work performed on the spaser; when $\pi < \Delta\varphi < 2\pi$, the spaser dipole moment supplies energy to the field. We are interested in the values of $E$ and the frequency detuning, at which $\Delta\varphi = \pi$, so that the exact loss compensation is achieved. The semi-classical theory predicts that there is a pumping threshold $D_{comp}$ (and respective $\gamma_p^{th(compensation)}$), below which the loss compensation is not possible [27,33]. This threshold is smaller than the generation threshold $D_{0th}^{(1)}$.

For $D_{comp} < D_0 < D_{0th}^{(1)}$, the results of the numerical solution of the system of equations describing the dynamics of the expected values of operators are shown in Fig. 1. In this figure we plot the phase difference, $\Delta\varphi$, as a function of the field amplitude, $E$, and frequency detuning, $\Delta_E = \omega_E - \omega_{SP}$. The dipole moment, $\mu_{NP}\langle \hat{a} \rangle + \mu_{TLS}\langle \hat{\sigma} \rangle$, arises due to the coherent spaser response to the external wave (for spontaneously excited plasmons the average dipole moment equals zero). One can see that depending on the values of $E$ and $\Delta_E$, the external field is either amplified or reduced. The exact loss compensation occurs for the values $E$ and $\Delta_E$ at the compensation curve (shown by the dashed line) at which $\Delta\varphi = \pi$. Note that this result is similar to the semi-classical prediction [27,33] shown in Fig. 2. However, unlike Refs. [27,33], in Fig. 1 spontaneous radiation is taken into account. In the low quantum regime, similar to the semiclassical results, the spaser dipole moment is of the same order of magnitude as in the regime of the developed generation.

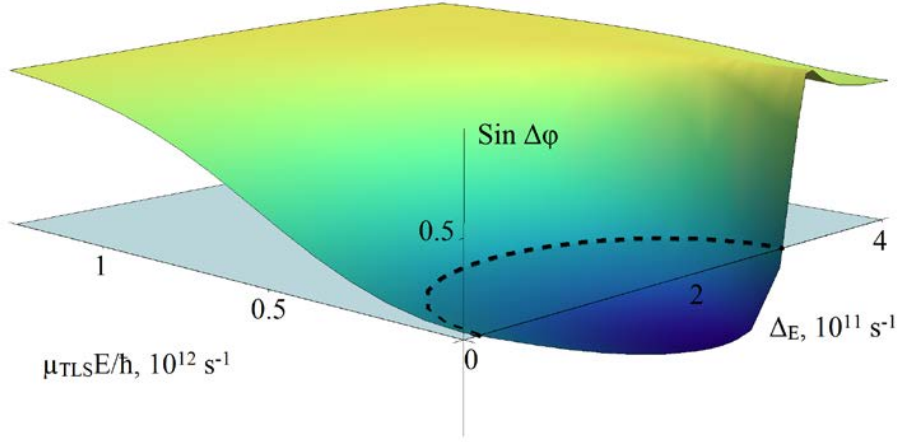

Fig. 1. The sine of the phase difference between oscillations of the spaser dipole moment and the external field as a function of the external field amplitude and the frequency detuning. The dashed line shows the compensation curve which corresponds to oscillations in antiphase with the external wave. The region bounded by this line corresponds to the amplification of the external field.

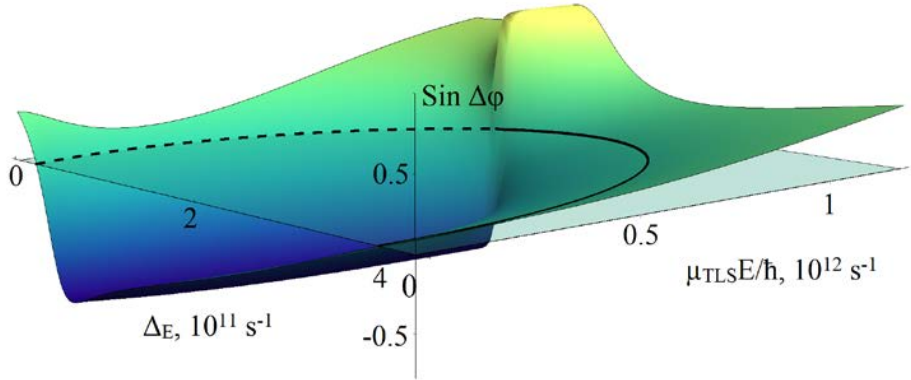

Fig. 2. The same as in Fig. 1 obtained in the semiclassical approximation [27,33]. The dashed line corresponds to oscillations in antiphase with the external wave, the solid line corresponds to in-phase oscillations.

Note, if spasers are used as loss compensating inclusions in metamaterials, then the amplitude of the electromagnetic wave travelling in the system will eventually be defined by its value on the compensation curve. Indeed, if the amplitude of the incident wave is above (below) the compensation curve, then during wave traveling, the energy is transferred to (from) the spasers until the wave amplitude reaches the compensation curve. Below we consider only these values of the external field [33,41].

In order to separate contributions of plasmons excited coherently and incoherently by spontaneous transitions, we note that both types of plasmons contribute to the averaged values of the operator $\hat{a}^+\hat{a}$. At the same time, induced transitions excite coherent oscillations of the dipole moment $\langle \hat{a} \rangle$. The energy of these oscillations is proportional to $|\langle \hat{a} \rangle|^2$. Therefore, we define the energy of coherently excited plasmons as $W_{coh} \sim \hbar \omega_{SP} |\langle \hat{a} \rangle|^2$ and the energy of incoherently plasmons as $W_{incoh} \sim \hbar \omega_{SP} \left( \langle \hat{a}^+ \hat{a} \rangle - |\langle \hat{a} \rangle|^2 \right)$ [9,10].

In Fig. 3, we compare quantities $|\langle \hat{a} \rangle|^2$ and $\langle \hat{a}^+ \hat{a} \rangle - |\langle \hat{a} \rangle|^2$ obtained on the compensation curve as functions of $\Delta_E$.

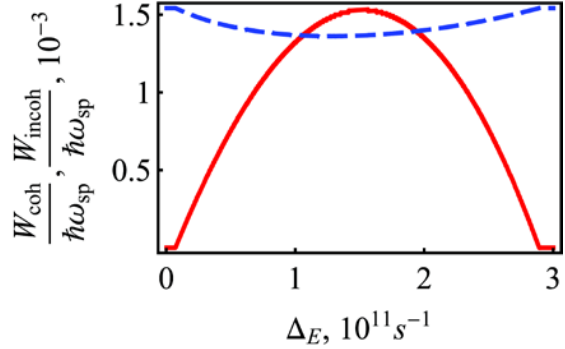

Fig. 3. The dependencies of the energy of the NP dipole moment, $W_{coh} \sim \hbar \omega_{SP} |\langle \hat{a} \rangle|^2$ (the solid line), and the energy of spontaneously excited plasmons, $W_{incoh} \sim \hbar \omega_{SP} \left( \langle \hat{a}^+ \hat{a} \rangle - |\langle \hat{a} \rangle|^2 \right)$ (the dashed line), calculated at the compensation curve near the first spasing threshold $\gamma_p / \gamma_p^{th(1)} \approx 1$, on the frequency detuning, $\Delta_E$. The parameters used in obtaining Fig. 3 are taken form Refs. [9,10]: $\gamma_a \sim 10^{14} s^{-1}$, $\gamma_\sigma \sim 10^{11} s^{-1}$, and $\Omega_R \sim 10^{13} s^{-1}$. These parameters correspond to a silver NP and a semiconducting active medium (e.g., InGaAs).

For further analysis, it is convenient to compare the results shown above with the results of the semiclassical theory in which operators are replaced by *c*-numbers [1,2,30]. This theory shows that spasing begins at the first pumping threshold, $D_{0th}^{(1)}$, below which there are no coherent oscillations, $|\langle \hat{a} \rangle|^2 = 0$, as shown by the solid line in Fig. 4a [1,2,36]. The principal difference between the driven spaser considered in this paper and an autonomous spaser is that the driven spaser ceases to be an auto-oscillating system and becomes a nonlinear oscillator interacting with an external harmonic force (the solid line in Fig. 4b).

Fig. 4. The dependencies of the energy of the NP dipole moment, $W_{coh} \sim \hbar\omega_{SP}|\langle \hat{a}\rangle|^2$ (the red solid line), and the energy of spontaneously excited plasmons, $W_{incoh} \sim \hbar\omega_{SP}(\langle \hat{a}^+\hat{a}\rangle - |\langle \hat{a}\rangle|^2)$ (the blue dashed line), on the pump rate characterized by $D_0$ for (*a*) an autonomous spaser and (*b*) a spaser driven by an external field.

In an autonomous spaser, the intensity of spontaneous generation, $W_{incoh}$, calculated in the low quantum regime is shown by the dashed line in Fig. 4a. We can see that the spaser is a low-quality laser in which the signal-to-noise ratio, $W_{coh}/W_{incoh}$, is very low. To achieve the value of the ratio greater than unity, one should employ the pumping $D_0 \gg D_{0th}^{(2)} > D_{0th}^{(1)}$, at which the number of plasmons $N_p$ is much greater than unity. Since in the low quantum regime it is assumed that $N_p \leq 1$, the developed theory only allows for an estimation of the threshold $D_{0th}^{(2)}$ which is done below.

For spasers, thermal losses far exceed radiation losses [38], therefore, the latter does not change the character of the attenuation and its impact on the constant $\gamma_a$ is negligible. The TLS relaxation rate, $\gamma_\sigma$, the pump rate, $\gamma_p$, and the expectation value of the operator $\hat{D}_0$, are related as [2]

$$\gamma_D = 2\gamma_\sigma + \gamma_p, \tag{7}$$

$$D_0 = (\gamma_p - 2\gamma_\sigma)/(\gamma_p + 2\gamma_\sigma). \tag{8}$$

Parameters $\gamma_a$, $\gamma_\sigma$, and $\Omega_R$ determine a semi-classical value of the pumping threshold $D_{th}^{(1)} = \gamma_a\gamma_\sigma/\Omega_R^2$. As follows from Eq. (8), the threshold pump rate, $\gamma_p^{th(1)}$ is

$$\gamma_p^{th(1)} = 2\gamma_\sigma(1 + D_{th}^{(1)})/(1 - D_{th}^{(1)}) \tag{9}$$

After finding the stationary solution of Eqs. (3)-(5), one can determine the number of coherently generated plasmons in the stationary regime [1,2,29]

$$n_c = \langle \hat{a}^\dagger\hat{a}\rangle_{coherent} = |a|^2 = (D_0 - D_{th}^{(1)})\gamma_D/4\gamma_a, \tag{10}$$

Let us now consider the spontaneous decay of the TLS excited state that leads to the radiationless excitation of a plasmon. We are interested in the case of large losses, for which $\gamma_a \gg \Omega_R, \gamma_\sigma, \gamma_D$. Below the first threshold, in the low quantum regime, the intensity of the field of plasmons excited due to spontaneous transitions of the TLS can be estimated as [37]

$$\langle \hat{a}^\dagger \hat{a} \rangle_{spontaneous} = \Omega_R^2 \gamma_a^{-2}(1+D_0)/2 \tag{11}$$

We may introduce the Purcell factor $F = \tilde{\gamma}_\sigma / \gamma_\sigma$ [44] which is the ratio of TLS decay rates in the presence of the NP, $\tilde{\gamma}_\sigma = \Omega_R^2/\gamma_a$ [30,37,45], and in free space, $\gamma_\sigma$. For a low-$Q$ cavity, $\gamma_a \gg \tilde{\gamma}_\sigma, \Omega_R$, and Eq. (11) can be rewritten as [30,37]

$$\langle \hat{a}^\dagger \hat{a} \rangle_{spontaneous} = F\gamma_\sigma \gamma_a^{-1}(1+D_0)/8 \tag{12}$$

Note that the Purcell factor is related to the first generation threshold [30]

$$F = 4\Omega_R^2 / \gamma_a \gamma_\sigma = 4/D_{0th}^{(1)}. \tag{13}$$

In the case under consideration, $D_{0th}^{(1)} = \gamma_a \gamma_\sigma / \Omega_R^2 \ll 1$, so that $F \gg 1$ (according to Refs. [9,10], $F \sim 100$) and $\gamma_p^{th(1)} \approx 2\gamma_\sigma$

$$\gamma_p^{th(1)} \approx 2\gamma_\sigma. \tag{14}$$

Let us find the value of the pump rate for which contributions of spontaneous and stimulated processes to the plasmon generation become comparable,

$$\langle \hat{a}^\dagger \hat{a} \rangle_{coherent} \sim \langle \hat{a}^\dagger \hat{a} \rangle_{spontaneous}. \tag{15}$$

Using Eqs. (10), (12), and Eq. (15) to estimate the value of $\gamma_p^{th(2)}$ we obtain

$$\gamma_p^{th(2)} \approx \frac{4\gamma_\sigma}{D_{th}^{(1)}} \sim F\gamma_p^{th(1)}. \tag{16}$$

$\gamma_p^{th(2)}$ defined by Eq. (16) was previously obtained in Ref. [9].

## 4. CONCLUSIONS

In the low quantum regime, we analyze the behavior of a spaser synchronized with an external electromagnetic wave. We demonstrate that the performance of a driven spaser is drastically different from an autonomous 3-D confined spaser. Unlike the latter, a driven spaser may mainly generate coherent surface plasmons even near or below the spasing threshold (see Fig. 3).

In the presence of an external field, a new parameter, $\Delta_E = \omega_E - \omega_{SP}$, arises. To a great extent, this parameter governs the behavior of the driven spaser. This can be seen in Fig. 1. When the detuning is near zero, the field on the compensation curve vanishes and the spaser is autonomous: the energy of the NP dipole moment, $W_{coh} \sim \hbar\omega_{SP}|\langle \hat{a} \rangle|^2$, nearly vanishes and most of the plasmons are incoherent. This is the case considered in Refs. [9,10]. When the detuning increases up to the critical value, the fraction of coherent plasmons increases. As a result, as shown in Fig. 3, for certain values of $\Delta_E$, we have $W_{coh} \geq W_{incoh}$, i.e., the energy of the excited dipole moment is comparable or greater than the energy of the spontaneous transitions.

In Ref. [46], spasers are built with a large number of active molecules $N$. To compare our further estimations with the experiment, we neglect the interaction between active particles simply setting $N \gg 1$.

Let us estimate the electric field amplitude on the compensation curve. The characteristic value of the matrix element of the TLS dipole moment is $\mu_{TLS} \sim 20\,\text{D}$; on the compensation curve $\mu_{TLS} E / \hbar \sim 10^{11}\,\text{s}^{-1}$, and then $E \sim 10^5\,\text{V/m}$. This value is by more than an order of magnitude smaller than the value of the field of electrical breakdown of air, $\sim 3 \cdot 10^6\,\text{V/m}$.

Now let us estimate the value of the pump rate necessary to reach the second generation threshold at which the ratio of signal-to-noise is about unity. This value is defined by Eq. (16). Keeping in mind that according to Eq. (14) for a low-$Q$ resonator $\gamma_p^{th(1)} \sim \gamma_\sigma$, we obtain $\gamma_p^{th(2)} \sim \Omega_R^2 / \gamma_a$. This relationship only weakly depends on the properties of an active medium. For typical values of $\Omega_R \sim 10^{13}\,s^{-1}$ and $\gamma_a \sim 10^{13}...10^{14}\,s^{-1}$, the second threshold pump rate is . To determine whether this value is attainable, we need to consider how it can be realized experimentally. If dye molecules are used as an active medium and pumping is realized by external electromagnetic radiation, then the electric field can be estimated to be [47], which is achievable in experiment as demonstrated in Ref. [46]. In the case of an active semiconductor medium pumped by an electric current, the situation is less clear. In Refs. [9,10], within the framework of the model of the classical dissipative transport, it was shown that the required current density is , which cannot be attained in experiment.

Our estimates show that for a spaser driven by an external optical wave loss compensation can be achieved for attainable pump power rates. Indeed, as Fig. 4 shows, . Realistic values of  can be realized for both laser pumping of dye molecules and current pumping of a semiconductor medium. At the same time, the signal-to-noise ratio can be reasonably high. This regime can be realized experimentally so that spasers may offer exciting prospects for a broad range of applications.

**Acknowledgments**

The authors would like to thank V. V. Klimov, Yu. E. Lozovik, N. M. Schelkachev, and S. I. Bozhevolnyi for helpful discussions. This work was partly supported by RFBR Grants Nos. 12-02-01093, 13-02-00407, 13-02-92660, by the Dynasty Foundation, by the NSF under Grant No. DMR-1312707, and by the PSC-CUNY award.